\documentclass[aps,prb,reprint,floatfix,superscriptaddress]{revtex4-2}
\usepackage{amsmath,amssymb,amsthm}
\usepackage{graphicx} 
\usepackage{float}
\usepackage[hidelinks]{hyperref}

\newtheorem{theorem}{Theorem}[section]

\theoremstyle{remark}
\newtheorem{remark}[theorem]{Remark}
\theoremstyle{definition}

\newcommand{\BZ}{\mathbb{T}^2}
\newcommand{\vol}{\mathrm{vol}}
\newcommand{\dd}{\mathrm{d}}

\begin{document}

\title{Hodge Topology of Semiclassical Transport: A Coordinate-Free Geometric Framework for the Anomalous Hall Effect and Non-Linear Berry Dipole}

\author{Zhi-Wei Wang}
\email{zhiweiwang.phy@gmail.com}
\affiliation{College of Physics, Jilin University, Changchun, 130012, People's Republic of China}
\affiliation{University of York, York YO10 5GH, United Kingdom}

\author{Samuel L. Braunstein}
\email{sam.braunstein@york.ac.uk}
\affiliation{University of York, York YO10 5GH, United Kingdom}

\date{\today}

\begin{abstract}
We establish a coordinate-free differential geometric framework for anomalous transport in topological bands, rooted in the Hodge-de Rham decomposition of the Brillouin zone. Standard geometric formulations of topological transport encounter mathematical singularities (Dirac strings) when attempting to utilize the quantum Berry connection in bands with non-zero Chern numbers. By applying the Hodge-de Rham decomposition directly to the Berry curvature 2-form, we isolate the quantized topological monopole flux from a globally defined, smooth geometric 1-form proxy potential, $\mathcal{A}$. We demonstrate that substituting this regularized exact potential into semiclassical transport integrals yields analytical advantages. For linear transverse transport, we show that our cohomological decomposition provides an exact geometric derivation of Haldane's insight via the mathematical co-area formula, partitioning the response into a continuous Fermi sea topological background and a localized Fermi surface geometric line integral. For non-linear transport, we demonstrate that this globally smooth proxy unifies the geometric description, reproducing the high numerical stability of standard scalar integration-by-parts techniques directly from its exact sector, while accommodating arbitrary Chern numbers. Furthermore, by enforcing the continuous Coulomb-Hodge gauge ($\delta \mathcal{A} = 0$) alongside vanishing harmonic holonomies over fundamental 1-cycles ($\oint_{\gamma_i} \mathcal{A} = 0$), we map the identity of the Hodge potential $\mathcal{A}$ to the Maximally Localized Wannier Function (MLWF) gauge for trivial bands, while serving as a non-singular computational proxy for topologically obstructed bands. Finally, we analytically demonstrate that solving the Hodge Laplacian for $\mathcal{A}$ zeroes the macroscopic Brillouin zone average (uniform $\mathbf{R}=0$ zero-mode) topological divergence, providing a mathematically consistent covariant formulation that matches the algorithmic robustness of standard analytical methods against discrete $\mathbf{k}$-grid noise.
\end{abstract}

\maketitle

\section{Introduction: Semiclassical Transport and Geometric Formulations}
\label{sec:intro}

The semiclassical formulation of electron dynamics provides the foundation for understanding macroscopic transport in crystalline solids \cite{sundaram1999, xiao2010}. Under a uniform in-plane DC electric field $\mathbf{E}$, the equations of motion for a wavepacket centered at position $\mathbf{r}$ and crystal momentum $\mathbf{k}$ are given by (using natural units $\hbar = 1$):
\begin{equation}
\dot{\mathbf{k}} = -e\mathbf{E}, \qquad \dot{\mathbf{r}} = \nabla_{\mathbf{k}} \varepsilon(\mathbf{k}) - \dot{\mathbf{k}} \times \mathbf{\Omega}(\mathbf{k}),
\end{equation}
where $\varepsilon(\mathbf{k})$ is the band dispersion and $\mathbf{\Omega}(\mathbf{k}) = \Omega(\mathbf{k})\hat{z}$ is the Berry curvature. 

In the presence of an electric field, the macroscopic transport properties are determined by the momentum-space probability distribution $f(\mathbf{k})$, which is driven out of equilibrium. Within standard Boltzmann kinetic theory utilizing the Relaxation Time Approximation (RTA), the steady-state distribution expands as $f = f_0 + f_1$, where $f_0(\varepsilon)$ is the equilibrium Fermi-Dirac distribution. The non-equilibrium excitation $f_1$ balances the linear drift with a phenomenological scattering rate $1/\tau$:
\begin{equation}
- \dot{\mathbf{k}} \cdot \nabla_{\mathbf{k}} f_0 \approx \frac{f_1}{\tau} \implies f_1 \approx e\tau \mathbf{E} \cdot \nabla_{\mathbf{k}} f_0 = e\tau E_a (\partial_a f_0).
\end{equation}

The total macroscopic charge current is obtained by integrating the velocity over the Brillouin zone torus $(M, g) = (\BZ, \delta_{ij})$:
\begin{equation} \label{eq:current_general}
\mathbf{J} = -e \int_{\BZ} \dot{\mathbf{r}} f(\mathbf{k}) \, \frac{\dd^2 k}{(2\pi)^2}.
\end{equation}

While the longitudinal current is dominated by the group velocity $\nabla_{\mathbf{k}} \varepsilon$, the anomalous velocity term ($- \dot{\mathbf{k}} \times \mathbf{\Omega}$) generates purely transverse Hall responses. The linear Anomalous Hall Effect (AHE) stems from integrating the equilibrium distribution $f_0$ with the Berry curvature, while next-order non-linear responses are driven by the non-equilibrium distribution $f_1$ \cite{haldane2004, sodemann2015, ma2019}. 

\textbf{Topological and Geometric Formulations:} In standard quantum mechanics, the Berry curvature originates as the curl of the quantum Berry connection $\mathbf{A}_{\text{Berry}} = i \langle u_n(\mathbf{k}) | \nabla_{\mathbf{k}} | u_n(\mathbf{k}) \rangle$. However, Chern's theorem mathematically guarantees that if $c_1 \neq 0$, a globally smooth, continuous definition of $\mathbf{A}_{\text{Berry}}$ cannot exist across the entire Brillouin zone. Any attempt to define it globally will spawn gauge singularities (Dirac strings). While standard computational methods successfully circumvent this by performing integrations over the single-valued scalar curvature $\Omega_z(\mathbf{k})$, relying purely on scalar calculus obscures the underlying differential-geometric phase space structure.

In this paper, we recast the transport framework within the coordinate-free differential geometry of forms on the Brillouin zone. We demonstrate that the Hodge-de Rham decomposition bypasses the topological obstruction of $\mathbf{A}_{\text{Berry}}$, providing a unified geometric framework for Fermi surface localization and non-linear topological transport.

\section{Cohomological Decomposition of the Berry Curvature}
\label{sec:hodge_Omega}

We map the semiclassical geometric ingredients into the language of
differential forms \cite{nakahara2003,frankel2011} on the closed 2D
Brillouin zone manifold $\BZ$. The Berry curvature is a globally
well-defined, manifestly gauge-invariant 2-form: $\Omega =
\Omega_z(\mathbf{k}) \dd k_x \wedge \dd k_y \in \Omega^2(\BZ)$. 

To bypass the topological obstruction of the quantum connection, we perform a Hodge decomposition directly on $\Omega$. The general Hodge decomposition of a $k$-form is $\omega = \dd \alpha + \delta \beta + \gamma$. Because $\Omega$ is a top-form (a 2-form on a 2-dimensional manifold), the space of 3-forms is trivially empty ($\Omega^3(\BZ) = 0$). Therefore, the co-exact generating form must be zero ($\beta = 0 \implies \delta \beta = 0$). As a top-form, $\Omega$ is also trivially closed ($\dd\Omega = 0$). The curvature therefore uniquely splits into a harmonic component and an exact component:
\begin{equation}\label{eq:Omega_hodge}
\Omega = \frac{2\pi c_1}{A_{\BZ}} \vol + \dd\mathcal{A},
\end{equation}
where $c_1 \in \mathbb{Z}$ is the first Chern number, $A_{\BZ}$ is the momentum-space area of the Brillouin zone, $\vol = \dd k_x \wedge \dd k_y$ is the standard area measure, and $\mathcal{A} \in \Omega^1(\BZ)$ is a globally defined 1-form representing a geometric proxy potential. 

\begin{remark}[The Smoothness of $\mathcal{A}$]
The advantage of Eq.~\eqref{eq:Omega_hodge} lies in its mathematical regularity. The fundamental reason the standard quantum Berry connection $\mathbf{A}_{\text{Berry}}$ develops Dirac string singularities is that it attempts to absorb the entirety of the topological monopole charge into an exact exterior derivative ($\dd \mathbf{A}_{\text{Berry}} = \Omega$). By explicitly projecting out the uniformly distributed topological flux fraction ($\frac{2\pi c_1}{A_{\BZ}} \vol$) into the harmonic sector, the remaining curvature is formally devoid of any net topological charge. Consequently, mathematical theory guarantees that the proxy potential $\mathcal{A}$ is a \textbf{globally smooth} 1-form everywhere on the Brillouin zone, agnostic to whether the band is topologically trivial or non-trivial.
\end{remark}

\section{Formalizing Haldane's Fermi Surface Identity}
\label{sec:streda}

The linear transverse anomalous Hall current is driven by the anomalous velocity evaluated over the equilibrium Fermi sea. Utilizing the differential form framework and substituting the current components from Eq.~\eqref{eq:current_general}, the transport integral is given by:
\begin{equation}\label{eq:Janom_form}
\mathcal{J}_{\mathrm{anom}}^i = -e^2 \epsilon^{ij} E_j \frac{1}{(2\pi)^2} \int_{\BZ} f_0 \Omega.
\end{equation}

In 2004, Haldane provided a physical insight regarding this anomalous Hall effect: while intrinsically a property computed over the entire filled Fermi sea, the non-quantized geometric response can be re-expressed as a property of the 1D Fermi surface \cite{haldane2004}. We will now demonstrate that substituting our Hodge decomposition formally proves this identity via geometric dimensional reduction.

Substituting the decomposition from Eq.~\eqref{eq:Omega_hodge} into the anomalous current integral:
\begin{equation}
\frac{1}{(2\pi)^2} \int_{\BZ} f_0 \Omega = \frac{c_1}{2\pi A_{\BZ}} \int_{\BZ} f_0 \, \vol + \frac{1}{(2\pi)^2} \int_{\BZ} f_0 \dd\mathcal{A}.
\end{equation}
The first term trivially evaluates to $\frac{c_1 \nu}{2\pi}$, where $\nu
= \frac{1}{A_{\BZ}} \int_{\BZ} f_0 \, \vol$ is the dimensionless
continuous band filling fraction (the bulk topological background). 
For a filled
band ($\nu = 1$), this recovers the TKNN quantised Hall
conductivity~\cite{tknn1982}.

To evaluate the exact term, we utilize the Leibniz rule for exterior derivatives, $\dd(f_0\mathcal{A}) = \dd f_0 \wedge \mathcal{A} + f_0 \dd\mathcal{A}$. Crucially, because our Hodge proxy potential $\mathcal{A}$ is guaranteed to be globally smooth (unlike the standard Berry connection), we can apply Stokes' theorem on the closed Brillouin zone manifold without boundary singularity corrections: $\int_{\BZ} \dd(f_0\mathcal{A}) = 0$. This yields:
\begin{equation} \label{eq:IBP_linear}
\int_{\BZ} f_0 \dd\mathcal{A} = - \int_{\BZ} \dd f_0 \wedge \mathcal{A} = \int_{\BZ} \mathcal{A} \wedge \dd f_0,
\end{equation}
where the final sign change stems from the anti-commutativity of the wedge product between 1-forms. 

\begin{theorem}[Geometric Anomalous Current]\label{thm:anom}
The real-space linear anomalous Hall current separates cleanly into a fractionalized topological bulk background and an exact geometric form isolated by the thermodynamic gradient:
\begin{equation}
\mathcal{J}_{\mathrm{anom}}^i = -e^2 \epsilon^{ij} E_j \left[ \frac{c_1 \nu}{2\pi} + \frac{1}{(2\pi)^2} \int_{\BZ} \mathcal{A} \wedge \dd f_0 \right].
\end{equation}
\end{theorem}

\textbf{Dimensional Reduction via Co-Area Formula:} The topological implications of Theorem \ref{thm:anom} become explicit at zero temperature ($T \to 0$). The equilibrium Fermi-Dirac distribution condenses into a Heaviside step function $f_0 = \Theta(\mu_F - \varepsilon)$, sharply defining the contiguous manifold of occupied states. Its exterior derivative localizes into a Dirac delta function pinned at the Fermi energy $\mu_F$:
\begin{equation}
\dd f_0 = \frac{\partial f_0}{\partial \varepsilon} \dd\varepsilon = -\delta(\varepsilon - \mu_F) \dd\varepsilon.
\end{equation}

Directly substituting this exact derivative into our geometric integral, the wedge product acts as a topological projector. Using the anti-commutativity of 1-forms ($\mathcal{A} \wedge \dd\varepsilon = - \dd\varepsilon \wedge \mathcal{A}$), we obtain:
\begin{equation}
\int_{\BZ} \mathcal{A} \wedge \dd f_0 = -\int_{\BZ} \delta(\varepsilon - \mu_F) \mathcal{A} \wedge \dd\varepsilon = \int_{\BZ} \delta(\varepsilon - \mu_F) \dd\varepsilon \wedge \mathcal{A}.
\end{equation}
Applying the mathematical co-area formula for differential forms, the exact normal differential $\dd\varepsilon$ foliates the 2D momentum-space integration measure over 1D constant-energy contours. Because the energy gradient $\dd\varepsilon$ defines the outward normal, the induced orientation matches the standard boundary orientation. The delta function thus collapses the integration onto the Fermi surface (FS) contour:
\begin{equation}
\int_{\BZ} \delta(\varepsilon - \mu_F) \dd\varepsilon \wedge \mathcal{A} = \int \dd E \, \delta(E - \mu_F) \oint_{\varepsilon=E} \mathcal{A} = \oint_{\text{FS}} \mathcal{A}.
\end{equation}

By absorbing the Dirac string singularities into the bulk $c_1$ term, the Hodge proxy $\mathcal{A}$ provides the exact mathematical framework to localize anomalous phase space phenomena to the 1D Fermi boundary.

\section{Universal Topological Robustness: Geometric Unification of the Berry Dipole}
\label{sec:limit}

The exact formalism yields a unified foundation when calculating next-order responses. Driven out of equilibrium by the RTA distribution $f_1 = e\tau E_a (\partial_a f_0)$, the non-linear anomalous Hall effect (NLAHE) evaluates to:
\begin{eqnarray}\label{eq:NLAHE}
\mathcal{J}_{\mathrm{anom}}^{\mathrm{NL}, i}
&=& -e^2 \epsilon^{ij} E_j \frac{1}{(2\pi)^2} \int_{\BZ} f_1 \Omega
\nonumber \\
&=& -e^3 \tau \epsilon^{ij} E_j E_a \frac{1}{(2\pi)^2} W_a,
\end{eqnarray}
where the non-linear transport tensor coefficient is defined as $W_a \equiv \int_{\BZ} (\partial_a f_0) \Omega_z \, \vol$. 

To bypass the computational grid noise inherent in taking discrete numerical momentum derivatives ($\mathbf{k}$-gradients) of the Berry curvature, standard numerical packages correctly invoke an analytical scalar integration-by-parts to evaluate the dipole $D_a \equiv \int_{\BZ} f_0 (\partial_a \Omega_z(\mathbf{k})) \, \vol$. By shifting the momentum derivative away from the noisy curvature and onto the smooth equilibrium distribution, the coefficient evaluates as:
\begin{equation} \label{eq:standard_IBP}
- \int_{\BZ} f_0 (\partial_a \Omega_z) \, \vol = \int_{\BZ} (\partial_a f_0) \Omega_z \, \vol \equiv W_a.
\end{equation}

\textbf{Geometric Unification via the Hodge Proxy Potential:} This standard computational methodology relies entirely on scalar calculus; because both the equilibrium distribution $f_0(\mathbf{k})$ and the gauge-invariant Berry curvature $\Omega_z(\mathbf{k})$ are smooth, single-valued periodic scalar functions on the Brillouin zone torus, the fundamental theorem of calculus guarantees the exact identity $\int \partial_a(f_0 \Omega_z) \vol = 0$ for any arbitrary Chern number. There is no mathematical contradiction in standard numerical methods.

However, to incorporate non-linear responses into a unified, coordinate-free differential geometric framework—one that manipulates exact exterior forms and bounds integration to boundary manifolds—we can deploy the globally smooth Hodge potential $\mathcal{A}$.

Substituting the decomposition $\Omega = \frac{2\pi c_1}{A_{\BZ}} \vol + \dd\mathcal{A}$ directly into the dipole tensor coefficient:
\begin{eqnarray}
    W_a &=& \int_{\BZ} (\partial_a f_0) \left( \frac{2\pi c_1}{A_{\BZ}} \vol + \dd\mathcal{A} \right) \nonumber \\
    &=& \frac{2\pi c_1}{A_{\BZ}} \int_{\BZ} (\partial_a f_0) \vol + \int_{\BZ} (\partial_a f_0) \dd\mathcal{A}.
\end{eqnarray}

First, we address the topological monopole term. Because the equilibrium distribution $f_0(\mathbf{k})$ is periodic on the closed Brillouin zone, the term $(\partial_a f_0) \vol$ constitutes an exact 2-form: $(\partial_a f_0)\vol = \dd(f_0 \epsilon_{ab} \dd k_b)$ (where $\epsilon_{ab}$ is the 2D Levi-Civita symbol). The integral of any exact form over a closed manifold identically vanishes by Stokes' theorem:
\begin{equation}
\int_{\BZ} (\partial_a f_0) \vol = \int_{\BZ} \dd(f_0 \epsilon_{ab} \dd k_b) = 0.
\end{equation}
Remarkably, the non-equilibrium thermodynamic gradient filters out the topological monopole background, demonstrating geometrically that the $c_1$ dependence drops out. 

The non-linear macroscopic integral therefore isolates to the exact geometric proxy term:
\begin{equation}
W_a = \int_{\BZ} (\partial_a f_0) \dd\mathcal{A}.
\end{equation}
Now, following the exact same Leibniz rule expansion detailed in Eq.~\eqref{eq:IBP_linear} ($\dd ( (\partial_a f_0)\mathcal{A} ) = \dd(\partial_a f_0) \wedge \mathcal{A} + (\partial_a f_0) \dd\mathcal{A}$), we arrive at:
\begin{equation} \label{eq:W_exact}
W_a = \int_{\BZ} \dd ( (\partial_a f_0)\mathcal{A} ) - \int_{\BZ} \dd(\partial_a f_0) \wedge \mathcal{A}.
\end{equation}
Because we mathematically guarantee that $\mathcal{A}$ is a globally smooth 1-form devoid of Dirac strings, Stokes' theorem applies to the entire Brillouin zone torus, forcing the first boundary integral to exactly zero. 

Thus, the Hodge decomposition analytically reproduces the non-linear transport tensor entirely within the exact exterior calculus sector:
\begin{equation}
W_a = - \int_{\BZ} \dd(\partial_a f_0) \wedge \mathcal{A}.
\end{equation}
This geometric derivation shows that by utilizing the globally defined proxy potential $\mathcal{A}$, the Hodge framework unifies both linear and non-linear transport as operations purely on exterior forms, accommodating any topological charge without obstruction.

\section{The Physical Identity of \texorpdfstring{$\mathcal{A}$}{A}: A Smooth Gauge Proxy}
\label{sec:quantum}

In defining our geometric proxy via $\dd\mathcal{A} = \Omega - \frac{2\pi c_1}{A_{\BZ}} \vol$, the 1-form $\mathcal{A}$ possesses a residual, standard exact gauge degree of freedom ($\mathcal{A} \to \mathcal{A} - \dd\theta$). According to the Hodge Decomposition Theorem on a closed manifold, specifying both the exterior derivative and codifferential uniquely fixes a differential form up to an arbitrary harmonic 1-form $\gamma$. To uniquely fix the identity of $\mathcal{A}$ on the manifold, we explicitly impose the continuous Coulomb-Hodge gauge via the codifferential ($\delta = -\star \dd \star$), mathematically requiring:
\begin{equation}
\delta \mathcal{A} = 0,
\end{equation}
while simultaneously explicitly defining its harmonic component to vanish (i.e., no uniform constant vector fields), yielding the global holonomy constraint over the fundamental non-contractible 1-cycles (the generating reciprocal lattice loops $\gamma_x, \gamma_y$ of the Brillouin zone torus):
\begin{equation}
\oint_{\gamma_i} \mathcal{A} = 0 \quad (\text{for } i \in \{x,y\}).
\end{equation}
These continuous mathematical constraints unveil the physical identity of $\mathcal{A}$ with respect to modern electronic structure theory. Physically, because the Zak phase (the holonomy of the connection) determines the spatial center of the Wannier function, explicitly enforcing $\oint_{\gamma_i} \mathcal{A} = 0$ fixes the Wannier centers exactly at the real-space origin.

\textbf{Case 1: Topologically Trivial Bands ($c_1 = 0$).}
When a band is globally trivial, the bulk monopole term drops out entirely, yielding $\dd \mathcal{A} = \Omega$. In this regime, the physical quantum Berry connection $\mathbf{A}_{\text{Berry}}$ is capable of being globally smooth, meaning we can mathematically equate $\mathcal{A} = \mathbf{A}_{\text{Berry}}$. 

In computational solid-state physics, constructing Maximally Localized Wannier Functions (MLWFs) \cite{marzari2012} necessitates fixing the phase gauge of the connection. Marzari and Vanderbilt proved that minimizing the spatial spread functional of Wannier orbitals in real space identically necessitates enforcing the momentum-space Coulomb gauge: $\nabla_{\mathbf{k}} \cdot \mathbf{A}_{\text{Berry}} = 0$. Because the divergence of a vector field maps exactly to the negative codifferential of a 1-form ($\delta \mathbf{A}_{\text{Berry}} = 0$), enforcing the Coulomb-Hodge gauge directly computes the MLWF phase-fixed connection.

\textbf{Case 2: Topologically Non-Trivial Bands ($c_1 \neq 0$).}
When a band is topological, MLWFs are obstructed; one cannot enforce a globally smooth gauge on $\mathbf{A}_{\text{Berry}}$. However, the proxy potential $\mathcal{A}$ bypasses this exact constraint. By subtracting the uniform monopole charge, $\mathcal{A}$ operates as the smooth mathematical counterpart to the MLWF connection. It encapsulates the transverse geometric phase information required for evaluating exact Fermi surface transport properties (as proven in Section \ref{sec:limit}), while evading the topological obstructions that plague raw quantum mechanical wavefunctions.

\section{Algorithmic Consistency and Mathematical Identity}
\label{sec:stability}

The exactness of this proxy potential translates to mathematical consistency with stable computational algorithms. By imposing the gauge constraint $\delta \mathcal{A} = 0$, substituting $\mathcal{A}$ into the exterior derivative decomposition directly yields the Hodge Laplace equation:
\begin{equation}
\Delta_H \mathcal{A} = (\delta \dd + \dd \delta) \mathcal{A} = \delta \dd \mathcal{A} = \delta \left(\Omega - \frac{2\pi c_1}{A_{\BZ}} \vol\right).
\end{equation}

Our covariant geometric framework mathematically models why evaluating non-linear dipoles fundamentally manages to reject discrete background divergences. Transforming the Hodge Laplace equation into its exact conjugate space---the discrete real-space Bravais lattice vectors $\mathbf{R} = (R_x, R_y)$---algebraically maps the momentum derivatives via $\partial_{k_j} \to i R_j$. The coupled gauge and curl equations become:
\begin{eqnarray}
i R_x \tilde{\mathcal{A}}_x(\mathbf{R}) + i R_y \tilde{\mathcal{A}}_y(\mathbf{R}) &=& 0, \quad \text{and} 
\nonumber \\
i R_x \tilde{\mathcal{A}}_y(\mathbf{R}) - i R_y \tilde{\mathcal{A}}_x(\mathbf{R}) &=& \tilde{\Omega}_z(\mathbf{R}) - \frac{2\pi c_1}{A_{\BZ}}\delta_{\mathbf{R},0}.
\end{eqnarray}

Because the uniform zero-mode ($\mathbf{R}=0$) component of the continuous Berry curvature is exactly its average topological value ($\tilde{\Omega}_z(0) = \frac{1}{A_{\BZ}} \int \Omega_z \vol = \frac{2\pi c_1}{A_{\BZ}}$), the right-hand side is zeroed out at the origin. By explicitly performing this topological subtraction, the formulation avoids a division-by-zero divergence, enforcing the zero-holonomy constraint $\tilde{\mathcal{A}}(0) = 0$. For $\mathbf{R} \neq 0$, substituting the gauge constraint $\tilde{\mathcal{A}}_y = - \frac{R_x}{R_y} \tilde{\mathcal{A}}_x$ decouples the variables, yielding the explicit Fourier components:
\begin{equation}
\tilde{\mathcal{A}}_x(\mathbf{R}) = i\frac{R_y}{|\mathbf{R}|^2}\tilde{\Omega}_z(\mathbf{R}), \quad \text{and} \quad \tilde{\mathcal{A}}_y(\mathbf{R}) = -i\frac{R_x}{|\mathbf{R}|^2}\tilde{\Omega}_z(\mathbf{R}).
\end{equation}

While evaluating the geometric potential ($\mathcal{A}$) mathematically scales the geometric curvature by a factor of $1/|\mathbf{R}|$, it is crucial to recognize that the final physical observable does not gain an artificial filtering advantage over the established scalar algorithm. In the full geometric formulation ($W_a = - \int \dd(\partial_a f_0) \wedge \mathcal{A}$), the extra momentum exterior derivative $\dd(\partial_a f_0)$ re-injects a multiplying $|\mathbf{R}|$ factor, exactly cancelling the scaling. 

This algebraic equivalence can be definitively proven via Stokes' theorem in momentum space. Because our Hodge framework mathematically guarantees that $\mathcal{A}$ is single-valued and smooth:
\begin{eqnarray}
&&- \int_{\BZ} \dd(\partial_a f_0) \wedge \mathcal{A} 
= \int_{\BZ} (\partial_a f_0) \,\dd\mathcal{A} \\
&=& \int_{\BZ} (\partial_a f_0) \left( \Omega - \frac{2\pi c_1}{A_{\BZ}}\,\vol \right) = \int_{\BZ} (\partial_a f_0) \Omega_z \, \vol, \nonumber
\end{eqnarray}
where the integral of the exact momentum derivative of the periodic function $f_0$ precisely annihilates the bulk constant. 

\begin{figure}[ht]
    \centering
    \includegraphics[width=\linewidth]{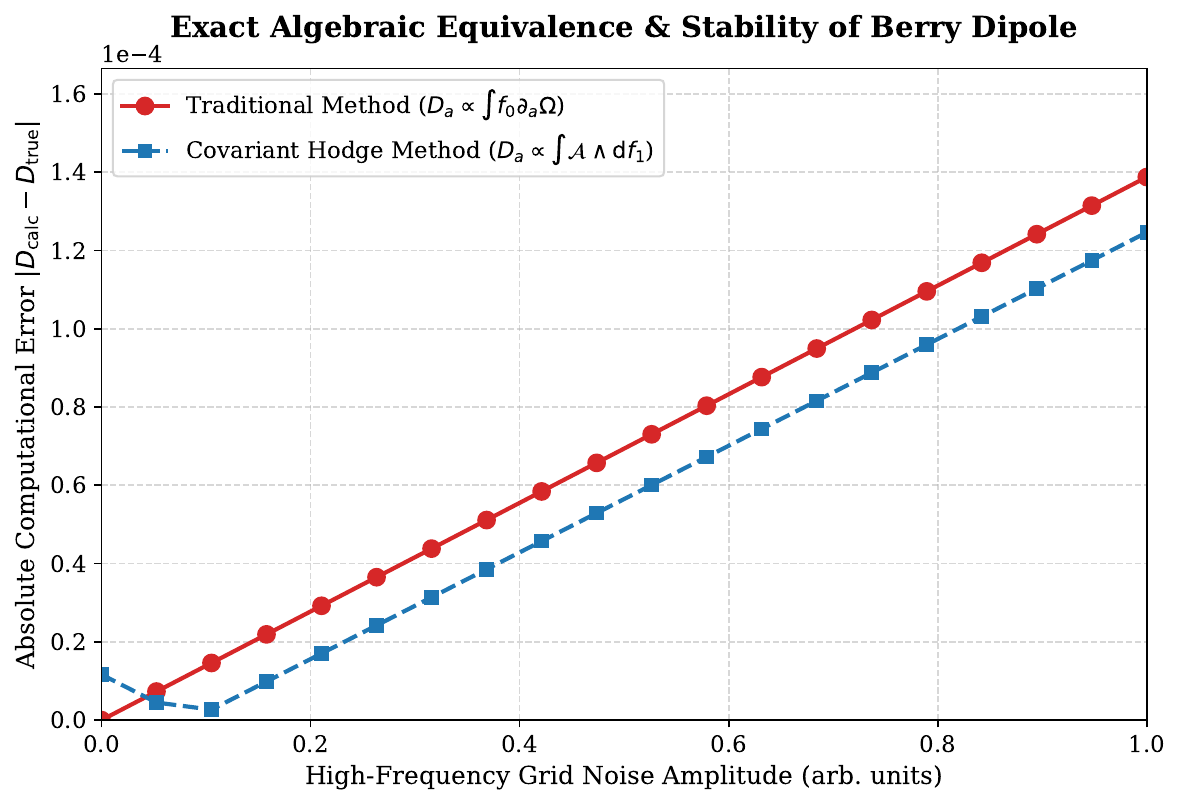}
    \caption{Numerical demonstration of the stability of extracting the non-linear Berry dipole ($D_x$) for a generic 2D tilted massive Dirac cone. Artificial high-frequency grid discretization noise is systematically injected into the underlying Berry curvature $\Omega(\mathbf{k})$. Both the standard scalar integration-by-parts method and the Covariant Hodge method ($\mathcal{A}$) demonstrate robust, bounded errors ($\sim 10^{-4}$). The Covariant Hodge framework formally explains this stability: operating within the geometric exact sector via $\mathcal{A}$ algebraically evaluates to the exact same noise-rejecting expression as standard scalar integration methods.}
    \label{fig:stability}
\end{figure}

As explicitly demonstrated in Figure~\ref{fig:stability}, applying the formal Hodge proxy integration (Eq.~\ref{eq:W_exact}) produces bounded errors identical to the standard analytical technique under severe artificial noise, because the methods algebraically compute the identical sum. Rather than competing with standard algorithms, our coordinate-free formulation provides the globally valid differential-geometric foundation that unifies and demonstrates their exact numerical stability, accommodating any arbitrary topological band structure.

\section{Outlook: 3D Weyl Topology}
\label{sec:outlook}

While this mathematical foundation models the 2D torus $\BZ$ where $\Omega$ is trivially closed ($\dd \Omega = 0$), extending this geometric analysis to the 3D Brillouin zone ($\mathbb{T}^3$) provides compelling avenues for future work. 

In 3D topological Weyl semimetals \cite{armitage2018}, the bands possess singular gap closing points. These Weyl nodes operate explicitly as momentum-space magnetic monopoles, enforcing distinct topological defects in the curvature tensor:
\begin{equation}
\dd\Omega = 2\pi \sum_n \chi_n \delta^{(3)}(\mathbf{k} - \mathbf{k}_n)\vol_3,
\end{equation}
where $\chi_n = \pm 1$ represents the quantized chirality of the node, and $\vol_3 = \dd k_x \wedge \dd k_y \wedge \dd k_z$. 

Because $\Omega$ is no longer universally closed in the 3D manifold, applying the full 3D Hodge decomposition uniquely spawns a non-trivial co-exact sector ($\Omega = \dd\mathcal{A} + \delta \mathcal{B} + \gamma$). The interplay of the exact proxy potential $\mathcal{A}$ and the co-exact geometric form $\mathcal{B}$ triggered by these monopoles offers a coordinate-free differential geometric pathway to formulate macroscopic non-equilibrium chiral anomalies \cite{fukushima2008} directly from topological flux distributions.

\end{document}